\documentclass[%
 reprint,
 amsmath,amssymb,
 aps,
]{revtex4-2}

\usepackage{graphicx}
\usepackage{dcolumn}
\usepackage{bm}
\usepackage{xcolor}
\usepackage{multirow}%
\usepackage{xcolor}%
\usepackage{hyperref}
\usepackage[normalem]{ulem}
\definecolor{olivegreenaccent3lighter60}{RGB}{215, 228, 189}
\definecolor{olivegreenaccent3darker25}{RGB}{119, 147, 60}
\definecolor{lavender}{RGB}{177, 119, 222}
\definecolor{redaccent2}{RGB}{192, 80, 77}
\definecolor{gold}{RGB}{255, 215, 0}
\newcommand{\filledcircle}[1]{\mbox{\textcolor{#1}{\circle*{5}}}}

\begin{document}

\preprint{APS/123-QED}

\title{Evaluating the transport properties of interface-type, analog memristors}

\author{Sahitya V. Vegesna}
 \email{sahitya.vegesna@uni-jena.de}
 \altaffiliation{One of the corresponding authors}
 \affiliation{Department of Quantum Detection, Leibniz Institute of Photonic Technology (IPHT), Albert-Einstein-Straße 9, 07745 Jena, Thuringia, Germany.}
 \affiliation{Institute for Solid State Physics, Friedrich Schiller University Jena, Helmholtzweg 3, 07743 Jena, Thuringia, Germany.}
 
\author{Venkata Rao Rayapati}%
 \email{rayapativenkat.crr@gmail.com}
 \altaffiliation{Contributing author}
 \affiliation{Department of Quantum Detection, Leibniz Institute of Photonic Technology (IPHT), Albert-Einstein-Straße 9, 07745 Jena, Thuringia, Germany.}%

\author{Heidemarie Schmidt}
 \email{heidemarie.schmidt@uni-jena.de}
 \altaffiliation{One of the corresponding authors}
 \affiliation{Department of Quantum Detection, Leibniz Institute of Photonic Technology (IPHT), Albert-Einstein-Straße 9, 07745 Jena, Thuringia, Germany.}
 \affiliation{Institute for Solid State Physics, Friedrich Schiller University Jena, Helmholtzweg 3, 07743 Jena, Thuringia, Germany.}

\date{\today}

\begin{abstract}
Interface-type, analog memristors have quite a reputation for real-time applications in edge sensorics, edge computing, and neuromorphic computing. The \textit{n}-type conducting \textit{BiFeO}\textsubscript{3} (\textit{BFO}) is such an interface-type, analog memristor which is also nonlinear and can therefore not only store, but also process data in the same memristor cell without data transfer between the data storage unit and the data processing unit. Here we present a physical memristor model which describes the hysteretic current-voltage curves of the \textit{BFO} memristor in the small and large current-voltage range. Extracted internal state variables are reconfigured by the ion drift in the two write branches and are determining the electron transport in the two read branches. Simulation of electronic circuits with the \textit{BFO} interface-type, analog memristors was not possible so far because previous physical memristor models have not captured the full range of internal state variables. We show quantitative agreement between modeled and experimental current-voltage curves exemplarily of three different \textit{BFO} memristors in the small and large current-voltage ranges. Extracted dynamic and static internal state variables in the two full write branches and in the two full read branches, respectively, can be used for simulating electronic circuits with \textit{BFO} memristors, e.g. in edge sensorics, edge computing, and neuromorphic computing.
\end{abstract}

\maketitle


\section{\label{sec1}Introduction}
The advent of the Internet of Things (\textit{IoT}) has transformed our technologically interconnected society, leading to a surge in the demand for computational resources. This increase in computational needs is intimately linked with a growth in energy consumption, a situation that is becoming increasingly untenable as we approach the limitations defined by Moore’s law. Consequently, the pursuit of more efficient computational methodologies has become imperative for the next phase of technological advancement. Research has pointed to innovative pathways such as the development of improved algorithms \cite{algo1,algo2,algo3} and the exploration of alternative memory elements \cite{10.1063/1.3153944,spagnolo_experimental_2022,aryana_interface_2021} that deviate from conventional semiconductor technology. Such alternative memory elements are one of the core enabler for novel artifical intelligence (\textit{AI}) accelerator hardware. \textit{AI} design needs to handle a lot of design corners, e.g. the development of physical models describing the new core enabler hardware. Within this context, memristors have emerged as a promising solution.

\begin{figure*}
  \centering
\includegraphics[scale=0.115]{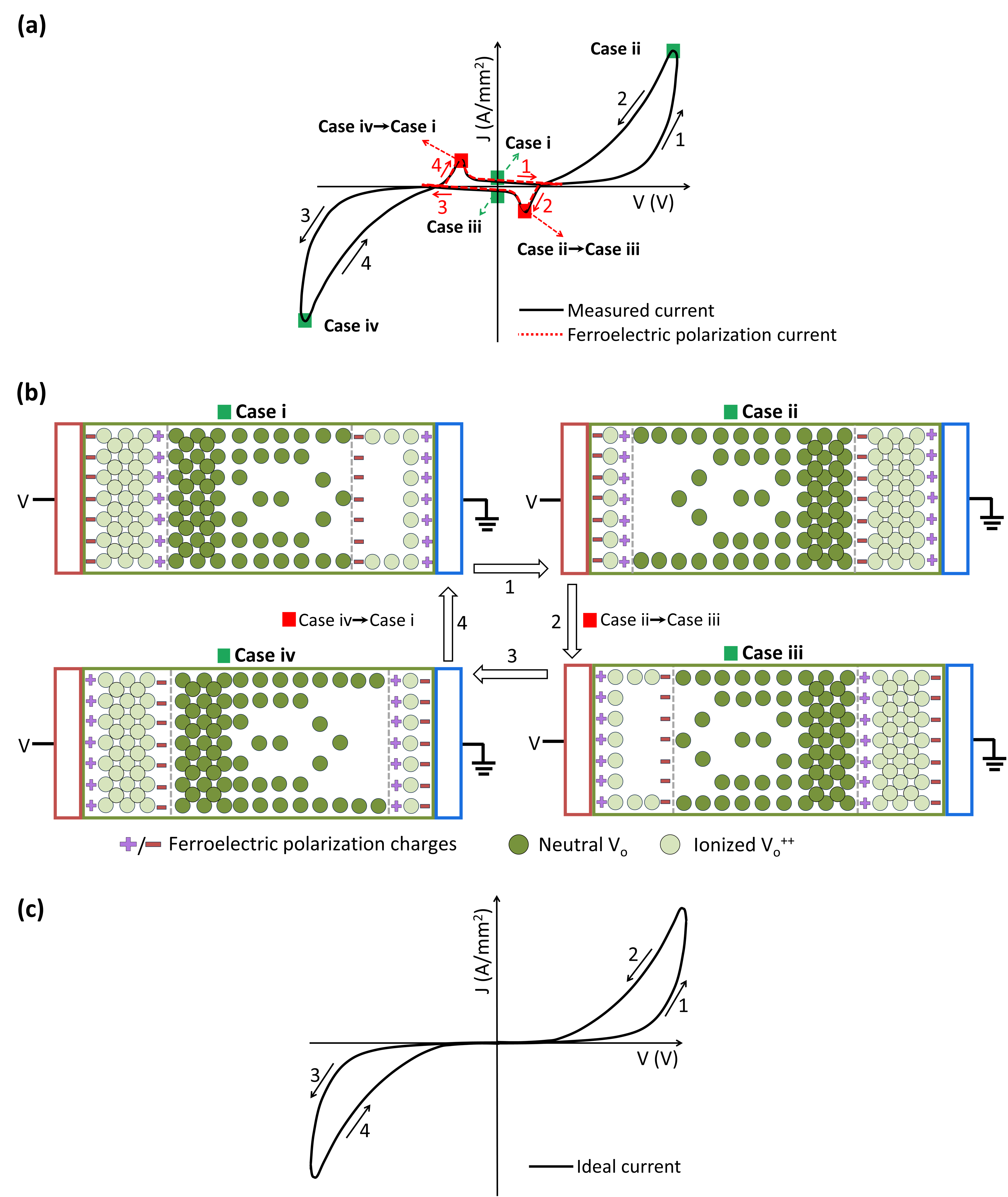}
\caption{\label{fig:Intro} (\textbf{a}) Schematics of a current density \(J\) - voltage \(V\) characteristic curve of a \textit{BFO} memristor before ferroelectric correction (black-solid line) and superimposed schematics of a current density \(J\) - voltage \(V\) characteristic curve of ferroelectric polarization current (red dotted line). (\textbf{b}) Distribution of ferroelectric polarization charge and of neutral and ionized oxygen vacancies in a \textit{BFO} memristor in different \textbf{cases} \textbf{i-iv}, where the voltage is sourced to the top electrode (\textit{TE}) (red-lined box) of \textit{BFO} memristor with bottom electrode (\textit{BE}) (green-lined box) being grounded: \textbf{Case i} corresponds to the end of Branch 4 and the start of branch 1 at zero applied bias \(V = 0\, \text{V}\). \textbf{Case ii} portrays the initiation of Branch 2 and the termination of Branch 1 with a maximum positive bias \(V = +V_{\text{max}}\, \text{V}\). \textbf{Case iii} represents the commencement of Branch 3 and the end of Branch 2 at zero bias \(V = 0\, \text{V}\). \textbf{Case iv} depicts the onset of Branch 4 and the end of Branch 3 under maximum negative bias \(V = -V_{\text{max}}\, \text{V}\). Distribution of polarization charges (\textbf{\textcolor{lavender}{+}/\textcolor{redaccent2}{-}}) and neutral oxygen vacancies $\,^{\filledcircle{olivegreenaccent3darker25}}$ ($V_{o}$), along with ionized oxygen vacancies $\,^{\filledcircle{olivegreenaccent3lighter60}}$ ($V_{o}^{+}$, $V_{o}^{++}$), is depicted from \textbf{Case i} to \textbf{Case iv} (green squares in (\textbf{a})). The extension of the depletion layer of the \textit{TE} and of the \textit{BE} with ionized oxygen vacancies and the width of the undepleted layer with neutral oxygen vacancies is not to be scaled. The boundary of the depletion region below the \textit{TE} and the boundary of the depletion region below the \textit{BE} is marked by dotted grey lines. There are only the ferroelectric polarization charges in the depletion region. The ferro-electric switching is indicated by a polarity change of polarization charges. The positive polarization charges (\textcolor{lavender}{lavender \textbf{+}}) and negative polarization charges (\textcolor{redaccent2}{red \textbf{-}}) are flipped when going from \textbf{Case iv} to \textbf{Case i} (red triangles on branch 4 in (\textbf{a})) and when going from \textbf{Case ii} to \textbf{Case iii} (red triangles on branch 2 in (\textbf{a})). (\textbf{c}) Schematics of an ideal current density \(J\) - voltage \(V\) characteristic curve of a \textit{BFO} memristor after ferroelectric correction (black-solid line). In this work, we present a physical model that describes ideal \(J\)-\(V\) curves after ferroelectric correction (solid line in (\textbf{c})).}
\end{figure*}

Memristors serve as pivotal elements with extensive applicability \cite{Du_2018,8965060,https://doi.org/10.1002/adma.202003437,8388838} in advancing the domains of \textit{AI} \cite{AI1,AI2}, neural networks \cite{li_efficient_2018,NN} and secure electronics \cite{SE_Nan_Heidi,SE2}. In particular, the Bismuth iron oxide \textit{BiFeO$_{3}$} (\textit{BFO}) has attracted attention as an important material for interface-type, analog, and non-linear memristors \cite{BFO1,BFO2}. \textit{BFO} thin films are intrinsically \textit{n}-type conducting due to the formation of oxygen vacancy donor ions during fabrication. The current-voltage characteristics of \textit{BFO} thin films with two metallically conducting electrodes may reveal hysteretic behaviour. Such \textit{BFO}-based structures are called \textit{BFO} memristors. The hysteretic current-voltage characteristics of \textit{BFO} memristors reveal four current-voltage branches, two full reconfigurable write branches with dynamic internal state variables and two full unaltered read branches with static internal state variables \cite{HS2024}. Oxygen vacancies in \textit{BFO} thin films with two metallically conducting electrodes only drift between the two metallically conducting electrodes in the two write branches and not in the two read branches in the electric field of the applied voltage \cite{you_bipolar_2014}. The redistribution of oxygen vacancies in the two write branches can be described by a rigid-point ion model \cite{Du_2018}. In \textit{BFO} memristors, the diffusion of oxygen vacancies due to a chemical or thermal gradient is suppressed by substitutional \textit{Ti$^{4+}$} ions trapping the oxygen vacancies. Both the quasi-Fermi level ($E_F (\text{BFO})$) and the conduction band minimum in \textit{BFO} ($E_C$) depend on the distribution of the oxygen vacancies and determine the internal state variables of the \textit{BFO} memristor. Because the distribution of oxygen vacancies does not change in the two read branches, the internal state variables in the two read branches are unaltered and static. Because the distribution of oxygen vacancies changes in the two write branches, the internal state variables in the two write branches are reconfigurable and dynamic. Whether the \textit{BFO} memristor is operated in the read or in the write mode solely depends on whether the oxygen vacancy distribution in the \textit{BFO} memristor remains unchanged or is continuously changed, respectively. \textit{BFO} memristors show a changing memristance in the two full write branches and no change of memristance in the two full read branches \cite{Du_2013}. Furthermore, due to the non-linear current voltage curves, the resistance of the \textit{BFO} memristor is continuously changed in the two full read branches. Self-rectification characteristics are often observed in memristors with reconfigurable Schottky barrier, which appear in both measured and modeled data, but the corresponding analysis is inadequate \cite{Ferro_BFO_PLD,Yarragolla2022}. E.g., the study by Chen et al. \cite{Ferro_BFO_PLD} models experimental hysteretic current-voltage curves of \textit{BFO} memristors only adequately at large voltages. And the study by Yarragolla et al. \cite{Yarragolla2022} models hysteretic current-voltage curves of \textit{BFO} memristors adequately at small and large voltages, but assumes change of Schottky barrier not only in the two full write branches but also in the two full read branches. This results in inconsistencies pertaining to the reconfiguration of the quasi-Fermi level and to the modulation of the Schottky barrier height of the two metallically conducting electrodes attached to the \textit{BFO} memristor. Model presented here correctly describes that the Schottky barrier height can be reconfigured only in the two full write branches and not in the two full read branches. For advancing the development of electronic circuits with \textit{BFO} memristors a clear understanding of the dependence of internal state parameters on the operation of \textit{BFO} memristors is necessary \cite{model_need}. The physical memristor model of hysteretic current-voltage curves of interface-type memristors has to account for the dominating ion drift in the two write branches and its influence on the internal state variables which determine the electron drift in the two read branches. Such a model will pave the way for novel electrical and electronic devices with \textit{BFO} memristors that merge data processing and storage in the same device cell, essentially combining memory and processing without data transfer between memory and processor. Building on this, we propose a physical memristor model where the internal state variables only change in the two complete write branches, leaving the read branches unaffected. This new approach accurately reproduces the hysteretic IV characteristics of \textit{n}-type conducting, interface-type memristors in both small and large bias ranges and aligns the tendencies of the reconfigurable Schottky barrier height with theoretical band alignment predictions.

Figure\,\ref{fig:Intro} offers an insightful perspective into the read and write operation of \textit{BFO} memristors. \textit{BFO} is a multiferroic material with ferroelectric ordering (ferroelectric Curie temperature \(T_c = 1083\, \text{K}\) \cite{ferroelectric_Neel_temperature}) and antiferromagnetic ordering (antiferromagnetic N\'eel temperature \(T_N = 643\, \text{K}\) \cite{P_Fischer_1980}). When a voltage is sourced to operate the \textit{BFO} memristor below the ferroelectric N\'eel temperature, spontaneous electric polarization can be reversed and a polarization switching current flows \cite{Jang_2018}. Resistive switching in \textit{BFO} thin films with a metallically conducting bottom electrode and a metallically conducting top electrode has been reported and attributed to ferroelectric polarization current  \cite{BFO1,ferroI_BFO_Ovac}. We observed another type of resistive switching in \textit{BFO} thin films and attributed it to a reconfigurable Schottky barrier height of the electrodes \cite{Du_2018,Tiangui_You_2014}. In this work, we focus on analyzing the hysteretic current-voltage curves of \textit{BFO} thin films revealing exclusively resistive switching due to reconfigurable Schottky barrier heights of the electrodes and revealing no ferroelectric switching. We call \textit{BFO} memristors with such a hysteretic current-voltage curve “memristors with an ideal current” response (Fig.\,\ref{fig:Intro}\,c). The ideal current response is obtained from experimental current-voltage hysteresis curves (Fig.\,\ref{fig:Intro}\,a) by subtracting the contribution of possible polarization currents due to ferroelectric switching from experimental hysteretic current-voltage curves \cite{Ferro_correc}. Presented physical memristor model describes an ideal current response and has been tested on ideal hysteretic current-voltage curves of three different \textit{BFO} thin films \cite{Tiangui_You_2014} which we obtained by subtracting the tiny contribution of polarization currents due to ferroelectric switching from experimental hysteretic current-voltage curves. 

First, we analyze the change of the internal state variables of \textit{BFO} thin films with ideal current (\(I\))-voltage (\(V\)) characteristics curves under the electric field in the two write branches and  relate it with the redistribution of oxygen vacancies in \textit{BFO} memristors. Here, the current density \(J\), \(J=I/A\), with \(A\) being the area of top electrode (\textit{TE}) of \textit{BFO} memristors is introduced, because that way extracted internal state parameters are independent of the area of the \textit{TE}. We show experimental current density-voltage curves in Fig.\,\ref{fig:Intro}\,a and ideal current density-voltage curves in Fig.\,\ref{fig:Intro}\,c. The current \(I\) flows when the voltage \(V\) is ramped. We use ideal \(J\)-\(V\) hysteresis loops to extract the internal state variables based on the simulation. We also present the unique aspects of oxygen vacancy drift in \textit{BFO} memristor interface switching with the metallic conducting \textit{TE} and the metallic conducting bottom electrode (\textit{BE}), the \textit{TE} and \textit{BE} form a Schottky contact on the \textit{BFO} thin film where they are attached. The extension of the depletion layer is indicated with a grey scattered line in Fig.\,\ref{fig:Intro}\,(b). Notably, the ideal \(J\)-\(V\) curve is divided into four branches: Branch 1 and Branch 3 act as write branches, allowing modification to the internal state of the memristor, while Branch 2 and Branch 4 serve as read branches for retrieving the internal state which have been written in Branch 1 and Branch 3, respectively. This partitioning is essential in comprehending the underlying operational principles of the \textit{BFO} memristor. Accompanying the \(J\)-\(V\) curve, visualization of the distribution of mobile oxygen vacancies (\(V_o, V^+_o, V^{++}_o\)) in these different branches under various biases is explicitly depicted in Fig.\,\ref{fig:Intro}\,(b). In particular, the figure portrays neutral oxygen vacancies (\(V_o\)) in the undepleted region and ionized oxygen vacancies in the depletion layer (\(V^+_o, V^{++}_o\)). Oxygen vacancies only drift in an electric field in Branch 1 and Branch 3. They drift in the direction of the electric field and are continuously redistributed after a threshold electric field strength. The final distribution of ionized oxygen vacancies depends on the voltage vs time ramping profile (\(v=v(t)\)) in branches 1 and 3. If the internal state parameter at the beginning of a write branch is the same, then always the same memristance curve (flux-charge) will be recorded for the same memristor. It does not matter how a given distribution of ionized oxygen vacancies is reached.

In our study, we investigate four distinct current-voltage branches 1-4. The four IV branches differ with respect to the distributions of polarization charges and oxygen vacancies in \textit{BFO} memristors. The analysis yields internal state variables of \textit{BFO} memristors in the four IV branches 1-4. Armed with these insights into the internal state variables, it will become feasible to design and develop electronic circuits with \textit{BFO} memristors that unify data processing and storage functionality in the same cell without data transfer \cite{Tiangui_You_2014}, while also successfully suppressing ferroelectric switching, thereby enhancing their applicability and performance. In more insulating conventional \textit{BFO} thin films scenarios, the ferroelectric current predominantly manifests within the insulating region \cite{ferroI_BFO_Ovac}, particularly in the area defined as the depletion region. It typically occurs when there is an incremental increase in the potential applied across the ferroelectric material. Contrary to this established behavior, our observations reveal an intriguing anomaly: the insulating region in our investigation is synonymous with the depletion region at the \textit{TE} and \textit{BE} which form a Schottky contact. Given that the potential within this region is inversely related to the applied potential, the ferroelectric current is found to emerge as the applied potential diminishes \cite{Ferro_BFO_PLD}. This unexpected phenomenon presents a novel aspect of ferroelectric current behavior, opening a new dimension in our understanding of the underlying mechanics and may carry substantial implications for the design and operation of multiferroic memristors with interface switching where ferroelectric switching and its contribution to the current-voltage characteristics need to be suppressed, e.g., by random alignment of ferroelectric polarization in polycrystalline multiferroic memristors \cite{Supress_Ferro}. The multifaceted insights provided by Fig.\,\ref{fig:Intro} serve as a vital framework for future exploration and application in the field of memristor technology, as well as in the realm of multiferroic materials and oxygen vacancies drifting under the electric field in the write operation.

\section{\label{MOd}Modeling}
\begin{figure*}
  \centering
\includegraphics[width=\textwidth]{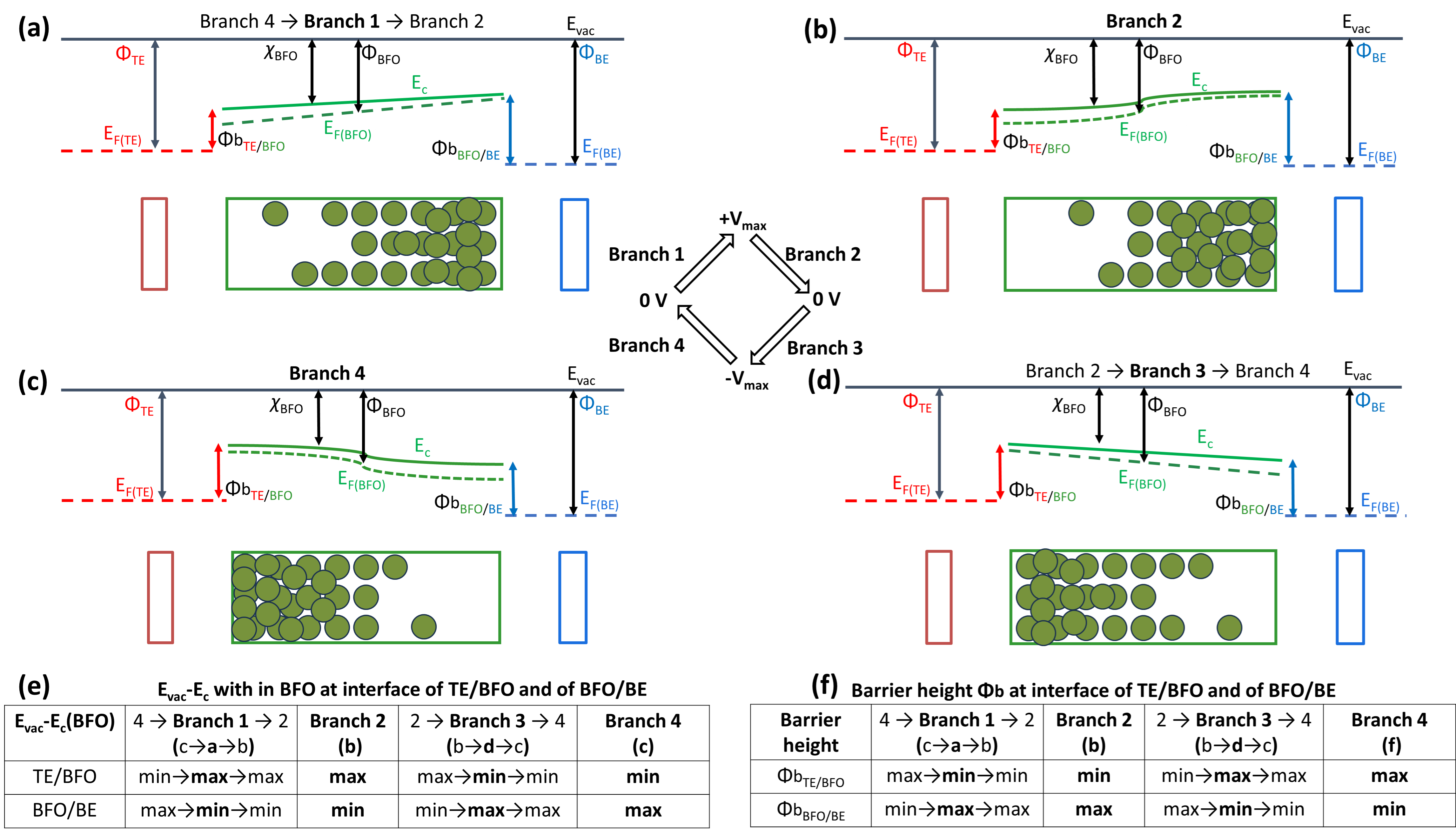}
\caption{\label{fig:bandE}Energy band diagrams and relevant parameters of studied \textit{TE/BFO/BE} structure where the \textit{TE}, \textit{BFO}, and \textit{BE} are not in contact. The quasi Fermi level in \textit{BFO} (\(E_F (\text{BFO})\)), the work functions of the \textit{TE} (\(\Phi_{\text{TE}}\)) and \textit{BE} (\(\Phi_{\text{BE}}\)), and the position dependent conduction band minimum of \textit{BFO} (\(E_c\)) are depicted with respect to the vacuum level ($E_{vac}$) in (\textbf{a}) An intermediate state in Branch 1, (\textbf{b}) Entire Branch 2, (\textbf{c}) An intermediate state in Branch 3, and (\textbf{d}) Entire Branch 4. The susceptibility of \textit{BFO} \(\chi_{\text{BFO}}\) is the difference between \(E_{vac}\) and \(E_c\). Work functions and Fermi level positions for the \textit{TE} and \textit{BE} are given as \(\Phi_{\text{TE}}\) and \(E_F (\textit{TE})\), and \(\Phi_{\text{BE}}\) and \(E_F (\textit{BE})\), respectively. The barrier heights at the \textit{TE/BFO} and \textit{BFO/BE} interfaces are denoted as \(\Phi{}b_{\text{TE/BFO}}\) and \(\Phi{}b_{\text{BFO/BE}}\), respectively. The color scheme is \textit{TE} (red) for the top electrode, \textit{BFO} (green) for the \textit{n}-type semiconductor, and \textit{BE} (blue) for the bottom electrode. Listed are (\textbf{e}) the variation of the energy difference \textbf{\(E_{vac} - E_c\)} within \textit{BFO} and (\textbf{f}) the barrier height \(\Phi_b\) at the \textit{TE/BFO} interface and at the \textit{BFO/BE} interface in dependence on the distribution of oxygen vacancies in Branch 4 \(\rightarrow\) Branch 1 \(\rightarrow\) Branch 2, in Branch 2, in Branch 2 \(\rightarrow\) Branch 3 \(\rightarrow\) Branch 4, and in Branch 4 by oxygen vacancy concentration.}
\end{figure*}

From now on we use area independent internal state variables which are extracted from the current density J vs. voltage curves. When examining the relationship between current density \(J\) - voltage \(V\) characteristics of \textit{BFO} memristors, we denote the layer structure of \textit{BFO} memristor as \textit{TE/BFO/BE}, with \textit{TE} representing the top electrode, \textit{BFO} representing the \textit{BFO} layer, and \textit{BE} representing the bottom electrode. Both junctions, the top electrode junction (\textit{TE/BFO}) and the bottom electrode junction (\textit{BFO/BE}) function as Schottky barriers (Fig.\,\ref{fig:Intro}). Schottky barriers (Fig.~\ref{fig:bandE}) play an instrumental role in the electronic properties of metal-semiconductor devices. In Fig.~\ref{fig:bandE}, we specify the \textit{TE} material as \textit{Au} and the \textit{BE} material as \textit{Pt/Ti}. Specifically, in the \textit{TE/BFO/BE} heterostructure, potential barriers emerge at the interfaces between the gold (\textit{Au}) top electrode (\textit{TE}) and the \textit{n}-type semiconductor \textit{BFO}, as well as between \textit{BFO} and the \textit{Pt/Ti} bottom electrode (\textit{BE}). When a metal, like \textit{Au} with a work function of \(\Phi_{\text{Au}} = 5.15\, \text{eV}\) \cite{AuPhi}, or \textit{Pt} with a work function of \(\Phi_{\text{Pt}} = 5.65\, \text{eV}\) \cite{PtPhi}, contacts the \textit{n}-type semiconductor \textit{BFO} with \(\chi_{\text{BFO}} = 3.3\, \text{eV}\) \cite{Chi_BFO}, electron migration occurs from the semiconductor to the metal until equilibrium is reached through Fermi level alignment. The oxygen vacancies are redistributed in the two full write branches of the hysteretic current-voltage curves of \textit{BFO} memristors. It is important to note that the alignment of the quasi-Fermi level \(E_F\) and of the conduction band minimum \(E_C\) depends on the distribution of free carriers and of oxygen vacancies, respectively. The dependence of the alignment of the conduction band minimum \(E_C\) has been related with the distribution of the oxygen vacancies and the partial density of states of oxygen vacancies around the conduction band minimum of \textit{BFO} \cite{DFT}. Because of reconfigured alignment of \(E_F\) and of \(E_C\), it is expected that also the internal state variables of the \textit{BFO} memristor depend on the oxygen vacancy distribution. In the presented Fig.~\ref{fig:bandE}, the intricate energy band dynamics of a \textit{TE/BFO/BE} heterostructure are meticulously showcased. Initial sections (\textbf{a}) to (\textbf{d}) provide comprehensive band diagrams depicting the transition and evolution of electronic states across different branches. Subsequently, (\textbf{e}) and (\textbf{f}) sections  offer insights into the influence of oxygen vacancy concentration on the energy difference within the \textit{BFO} at the interfaces and the barrier height parameters. This visualization is paramount in comprehending the correlation between susceptibility of \textit{BFO}, work functions, Fermi levels, and barrier heights. The color coding further simplifies the understanding, wherein the top electrode (\textit{TE}), the \textit{n}-type semiconductor (\textit{BFO}), and the \textit{BE} are represented in shades of red, green, and blue respectively. The figure’s comprehensive representation is pivotal for grasping the advanced principles and nuances of \textit{TE/BFO/BE} heterostructures, presenting it as a cornerstone for researchers and experts in the domain.

The Schottky barrier’s magnitude, \(\Phi{}b\), is determined by the energy difference between the conduction band minimum of \textit{BFO} ($E_C$) and the metal’s work function ($\Phi_{TE}$ or $\Phi_{BE}$). Interestingly, the total density of states at the conduction band minimum of \textit{BFO} ($E_C$) is only determined by the partial density of states of oxygen and iron, and not by the partial density of states of bismuth \cite{DFT}. Missing oxygen atoms will reduce the total density of states at the conduction band minimum of \textit{BFO} ($E_C$). Therefore, in comparison to stoichiometric \textit{BFO} without oxygen vacancies, for stoichiometric \textit{BFO} with oxygen vacancies, the conduction band minimum is shifted upwards and increases on the energy scale with increasing concentration of oxygen vacancies. The electron affinity is reduced correspondingly. Figures.~\ref{fig:bandE}(a-d) demonstrate the variation in the energy difference between the vacuum level ($E_{vac}$) and the conduction band minimum ($E_C$), as well as the energy difference between the conduction band minimum ($E_C$) and the metal's Fermi level ($E_F(TE)$ or $E_F(BE)$) for different oxygen vacancy distributions (branches 1-4), respectively. Qualitative values for these energy differences provided in Fig.~\ref{fig:bandE}(e) shows the energy difference between $E_{vac}$ and $E_C$, and in Fig.~\ref{fig:bandE}(f) presents the energy difference between $E_C$ and $E_F(TE)$ or $E_F(BE)$. The barrier heights at the interfaces of the \textit{Au} top electrode/\textit{BFO} and \textit{BFO/P/Ti} bottom electrode are denoted as $\Phi_b(TE/BFO)$ and $\Phi_b(BFO/BE)$, respectively. Because oxygen vacancies are intrinsic donors in \textit{BFO}, when being redistributed they consequentially modify the alignment of the quasi-Fermi level in \textit{n-BFO} (Fig.~\ref{fig:bandE}a-d). Therefore, the energy difference between the conduction band edge and the Fermi level in \textit{BFO}, labeled as \(E_C - E_F\), is regulated by the oxygen vacancy distribution. In essence, the Schottky barriers at the \textit{Au/BFO} and \textit{BFO/Pt/Ti} interfaces significantly impact the electronic properties of the heterostructure. Their heights are not only contingent upon the materials’ work functions but are also affected by internal factors like the oxygen vacancy distribution and the donor-quasi Fermi level concentrations in \textit{BFO}. Comprehensive understanding of these potential barriers is pivotal for predicting and enhancing the performance of devices stemming from such \textit{TE/BFO/TE} layer structures.

\begin{figure}
  \centering
\includegraphics[scale=0.1]{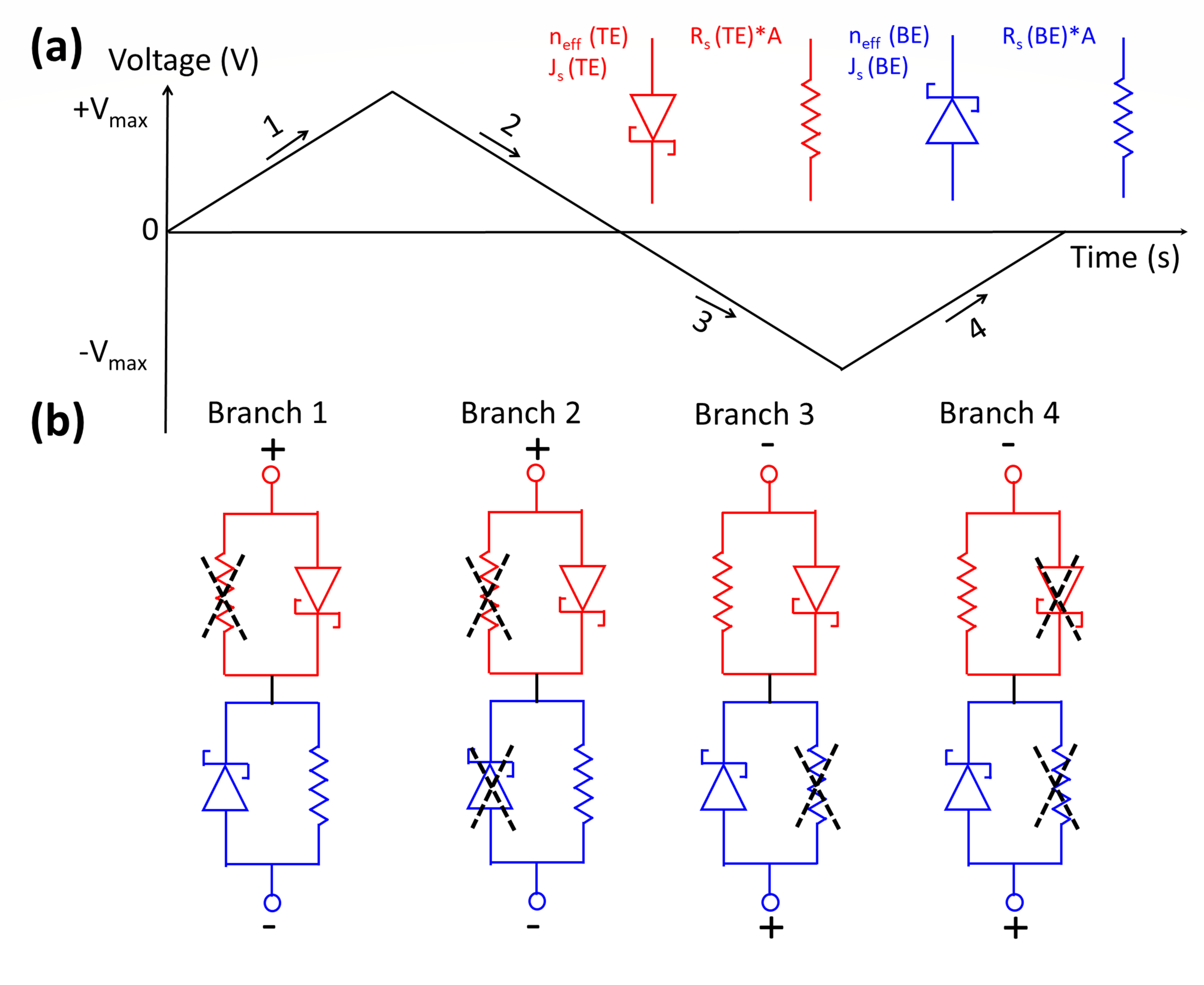}
\caption{\label{fig:Circuit} Illustration of the linear voltage sweep and its corresponding equivalent circuit for the four distinct branches: Branch 1 (0V to +\(V_{\text{max}}\)), Branch 2 (+\(V_{\text{max}}\) to 0\,V), Branch 3 (0\,V to -\(V_{\text{max}}\)), and Branch 4 (-\(V_{\text{max}}\) to 0\,V). The equivalent circuit includes a description of two Schottky diodes with ideality factor (\(n_{\text{eff}}\)) and reverse saturation current ($J_s$), along with a parallel leakage resistor ($R_s$). The \textit{TE/BFO} and \textit{BFO/BE} include the interface between \textit{BFO} and the top electrode and between \textit{BFO} and the bottom electrode, respectively. Black crossed-over lines within the diagram represent an insignificant circuit element for the specific branch under consideration. e.g, resistor in the connection \textit{TE/BFO} is an insignificant circuit element in branch 1.}
\end{figure}

Fig. \ref{fig:Circuit}(a) shows an example a linear voltage -time profile. The voltage-time profile determines the dynamics of internal parameter changes in the two full write branches. However, it does not influence the internal parameters in the two full read branches with their unique relation between applied voltage and applied current which does not depend on the charge flown and the history of the voltage during read operation. Applied voltage sweep is classified into four current-density branches, 1 (Branch 1, 0 to +\(V_{\text{max}}\)), 2 (Branch 2, +\(V_{\text{max}}\) to 0), 3 (Branch 3, 0 to -\(V_{\text{max}}\)) and 4  (Branch 4, -\(V_{\text{max}}\) to 0). The corresponding equivalent circuits are shown in Fig. \ref{fig:Circuit}(b) and used to model measured current density – voltage curves of three \textit{BFO} memristors [Fig. 2 in \cite{Tiangui_You_2014}].

 The corresponding equivalent circuits are shown in Fig.~\ref{fig:Circuit}(b) and used to model measured current density – voltage curves of three \textit{BFO} memristors [Fig.~\ref{fig:bandE} in \citet{Tiangui_You_2014}]. If, for a given depletion layer, the leakage resistance $R_S*A$ is very large, the leakage resistance can be neglected in the analysis of the corresponding Schottky barrier. Current flow over the leakage resistance $R_S*A$ in the depletion layer of the top electrode (branch 2 in Fig.~\ref{fig:Circuit}(b)) or over the leakage resistance $R_S*A$ in the depletion layer of the bottom electrode (branch 4 in Fig.~\ref{fig:Circuit}(b)) can be neglected if the concentration of oxygen vacancies is small. The extracted leakage resistance $R_S*A$ in the depletion layer with large concentration of oxygen vacancies, i.e., in the depletion layer of the bottom electrode (branch 2 in Fig.~\ref{fig:Circuit}(b)) and in the depletion layer of the top electrode (branch 4 in Fig.~\ref{fig:Circuit}(b)) are listed in Tab. I and range between a few to hundreds of $k\Omega\cdot \text{mm}^2$. Also, the Schottky barrier depends on the concentration of oxygen vacancies \cite{Vo_resistance,NOGUCHI20081006}. The prominent physical parameters describing the Schottky barriers are the reverse saturation current density ($J_s$) and the ideality factor ($n$). These parameters are integrated into the Schottky barrier equation, $J_s \propto e^{-\frac{\phi_{B}}{k_{B}T}}$, as follows:

\begin{equation}
 \label{eq:V_schot}
V_{i} = n\frac{ k_B T}{q} \ln \left( \frac{J}{J_s} + 1 \right),
\end{equation}
where $\phi_{B}$, \( J \), \(J_{\text{s}}\), \( q \), \( V_{i} \), \( n \), \( k_B \), and \( T \) are barrier potential, current density, reverse saturation current density, the charge of an electron, voltage drop across the barrier, ideality factor, Boltzmann constant, and temperature. The reverse saturation current density (\(J_{\text{s}}\)) provides insight into the minority carrier activities, while the ideality factor (\textit{n}) quantifies the divergence of the Schottky diode's real behavior from the ideal Schottky diode. 

In the present study, we also address the influence of the position dependent quasi-Fermi energy \(E_F\) in the \textit{n}-type \textit{BFO} layer as demonstrated in the energy band diagrams of Fig.~\ref{fig:bandE}, on the ideality factor \(n\). It has been observed that \(n\) is contingent upon the recombination mechanisms and remains constant provided that the energy difference \(E_c - E_F\) between the conduction band minimum and the donor quasi-Fermi energy within the \textit{BFO} remains uniform throughout. However, as the energy diagrams in Fig.~\ref{fig:bandE} parts (a) and (d) indicate, the presence of oxygen vacancy drift introduces variability in \(E_c - E_F\) at the interface of the \textit{TE/BFO/BE} layer structure. This variation and also the varitaion of \(E_{F(BFO)}\) in (b) and (c) necessitates a revision of the conventional ideality factor \(n\) to incorporate a dependency on the applied voltage, as the band alignment changes dynamically fluctuates due to the dynamic drift of oxygen vacancies. Consequently, a modification to the ideality factor is proposed as a function of voltage, allowing for a more nuanced understanding of the Schottky barrier behavior in the presence of non-uniform oxygen vacancy distribution in the \textit{BFO} layer and resulting non-uniform alignment of the donor-quasi Fermi level \(E_{F(BFO)}\) (Fig.~\ref{fig:bandE}).

Instead of relating the ideality factor with the total applied voltage across the \textit{TE/BFO/BE} structure, we introduce a current density (\( J \)) dependent effective ideality factor, denoted as \( n_{\text{eff}} \). This effective ideality factor accounts for the voltage drop across the specific contact under investigation. The relationship between the effective ideality factor and the applied voltage \( V_i \) and current density \( J \) is given as follows:
\begin{subequations}
\label{eq:n_eff}
\begin{align}
n_{\text{eff}} &= n  + (k \times V_i) \label{eq:n_eff0} \\
\intertext{where}
n_{\text{eff}} &= n \left(1 + k \frac{k_B T}{q} \ln\left(\frac{J}{J_s} + 1\right)\right) \label{eq:n_eff1}
\end{align}
\end{subequations}
Here, \( n \) is the conventional ideality factor from Eq.~\ref{eq:V_schot} and \( k \) [$V^{-1}$] is the proportionality constant. This reformulation offers a more targeted description of the device behavior, thereby enhancing the precision of the modeling and analysis specific to the contact configuration of interest. This novel approach paves the way for more accurate simulations and optimizations in various applications where the understanding of contact-specific behavior is crucial.

The \textit{TE/BFO} junction is modeled as a forward-biased Schottky diode (\textit{FB}), and the \textit{BFO/BE} junction as a reverse-biased Schottky diode (\textit{RB}), under conditions where the \textit{TE} is positively biased and the \textit{BE} is grounded \textit{(Fig. \ref{fig:Circuit}(b), Branch 1 and Branch 2)}. Conversely, when the \textit{TE} is negatively biased and the \textit{BE} is grounded, the \textit{TE/BFO} junction is represented by a reverse-biased(\textit{RB}) Schottky diode and the \textit{BFO/BE} junction by a forward-biased (\textit{FB}) Schottky diode \textit{(Fig. \ref{fig:Circuit}(b), Branch 3 and Branch 4)}.

\begin{figure*}
  \centering
\includegraphics[width=\textwidth]{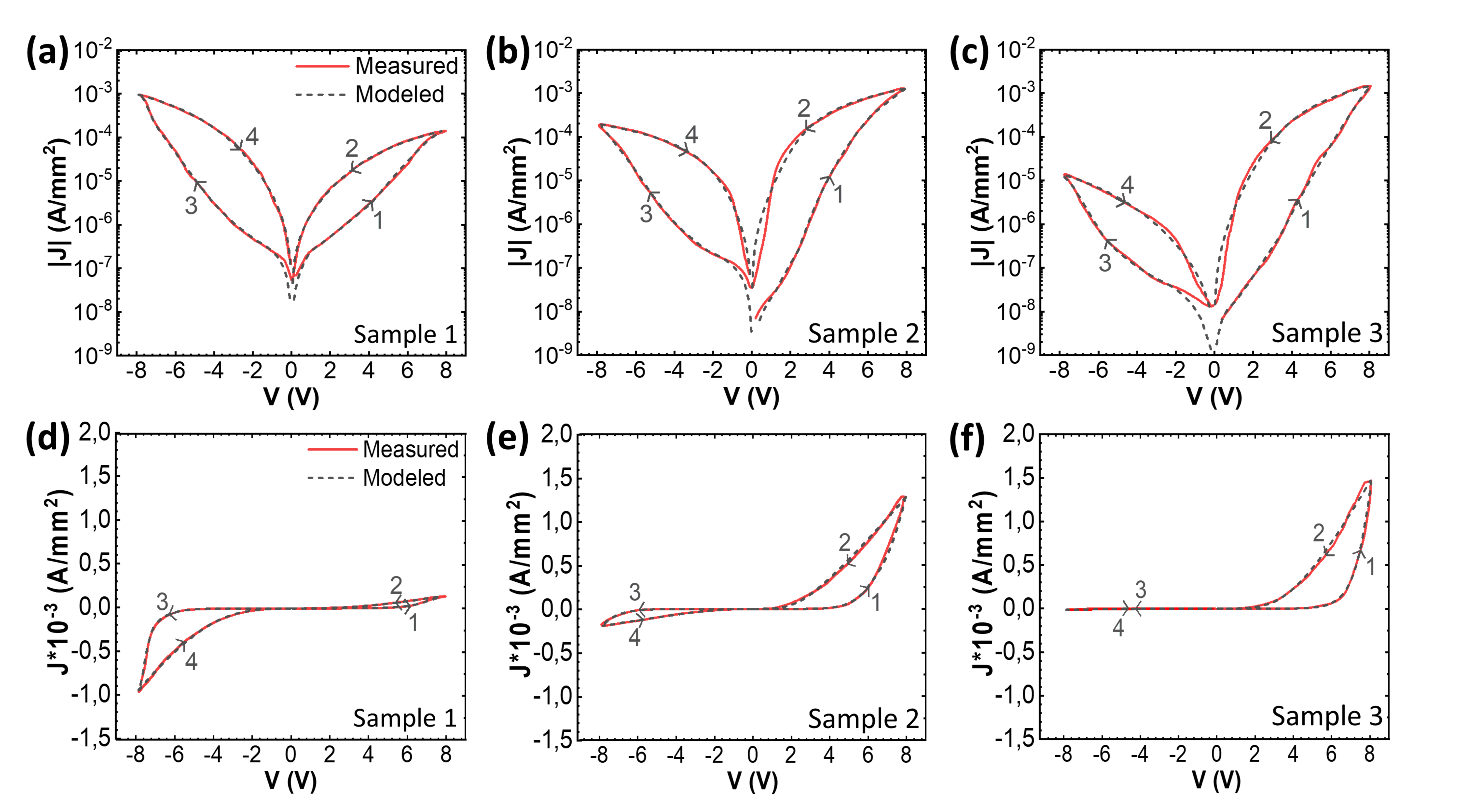}
\caption{\label{fig:EnM} Measured (red solid lines) and modeled (gray dashed lines) current density - voltage (J-V) characteristics shown in solid red lines of Samples 1-3 from T. You et al. \cite{Tiangui_You_2014} where (a,d) are Sample 1, (b,e) are Sample 2, and (c,f) are Sample 3. J in (a-c) is in log scale and (d-f) is in linear scale. Switching directions of the samples are 1 (Branch 1) $\rightarrow$ 2 (Branch 2) $\rightarrow$ 3 (Branch 3) $\rightarrow$ 4 (Branch 4). (1, 3) and (2, 4) correspond to WRITE and READ branches, respectively.\\}
\end{figure*}

Fig. \ref{fig:EnM} shows double hysteresis switching characteristics of \textit{BFO} memristor samples. The measured I-V data was taken from T. You et al. \cite{Tiangui_You_2014}. \textit{BFO} memristor shows an interface switching, the physical model parameters presented do not depend on the area of \textit{TE} and \textit{BE}. Therefore, we took the area (A =4.53 $\times$ 10$^{-2}$ mm$^2$) of \textit{TE} to model the current density J (J=I/A) - voltage (V) (red solid lines) characteristics and extract the physical model parameters in terms of J and R.A. The branches corresponding to WRITE branches are 1 \& 3 and READ branches are 2 \& 4. WRITE branches are characterized by the application of high-voltage ramping and the memristor is switched from one stable state (high resistance state) to the other (low resistance state). By applying opposite polarity, the resistance state (low resistance state) of READ branches (2,4) can be reversed. A circuit model is developed with a set of Eq's.~\ref{eq:Branchall} to simulate the effect of different distributions on the oxygen vacancies on the switching characteristics of \textit{BFO} samples. In the following, we apply presented simulation model to the current density-voltage hysteresis loops of three different \textit{BFO} memristors. We call \textit{J-V} curves with a hysteresis in the first and third \textit{J-V} quadrant double \textit{J-V} hysteresis. Those \textit{BFO} memristors with double J-V hysteresis have been fabricated and investigated in a previous work \cite{Tiangui_You_2014} to demonstrate the reconfigurability of \textit{BFO} memristors into the 16 Boolean logic gates. This was an important step towards data processing and storage in the same memristor device without data transfer. However, only now one can estimate the power consumption, speed, size and cost of electronic circuits with \textit{BFO} memristors using the presented physical model for electronic design automation simulations. The experimental current density-voltage hysteresis loops of three the different \textit{BFO} memristors presented only in Ref.\,\cite{Tiangui_You_2014} contain small contributions from the ferroelectric polarization current density. We removed the ferroelectric polarization current density and obtained an ideal current density (Fig.\,\ref{fig:Intro}). In the following we report in detail on the extraction of the internal state variables (Fig. \ref{fig:Circuit}b) from the double J-V hysteresis with, branches 1-4 (Fig. \ref{fig:Intro}c).

A comprehensive investigation of the double J-V hysteresis (Fig. \ref{fig:EnM}) reveals two prominent resistance states, necessitating two distinct input signals for transition between them, as evidenced by the experimental data in Ref. \cite{Tiangui_You_2014}. The measured \textit{J-V} characteristic curves of sample 1, sample 2, and sample 3 from You et al. \cite{Tiangui_You_2014} are depicted in Fig.\,\ref{fig:EnM}(a-c) on a logarithmic scale and in Fig.\,\ref{fig:EnM}(d-f) on a linear scale, after the subtraction of ferroelectric current density if necessary. Actually, the ferroelectric contribution of sample 1 and sample 3 is negligible and only the J-V curve of sample 2 shows a contribution from the ferroelectric polarization current in the small voltage range. We provide an ideal current density-voltage (\textit{J/V}) data set, derived from the data as initially reported in You et al. \cite{Tiangui_You_2014}. Branch labels, 1-4, have been included in both types of representations. The behavior of the memristor can be partitioned into two diodes, specifically, the forward bias (\textit{FB}) diode and the reverse bias (\textit{RB}) diode, which is expressed by the following equation:
\begin{widetext}
\begin{subequations}
\label{eq:Branchall}
\begin{flalign}
\text{All Branches:} & \nonumber \\
n_{\text{RB}}^{\text{eff}} &= \left( n_{RB} + n_{RB} k_{RB} \frac{k_{B}T}{q} \ln\left( \frac{-J_{D_{RB}}}{J_{S_{RB}}[-J]}+1 \right)\right), \label{eq:Branchalla} \\
  J_{D_{RB}} &= J_{S_{RB}}[J] \left( e^{\Delta} - 1 \right) \,\,\, where \,\,\, \Delta \,\,\, = \frac{-qV_{RB}}{n_{\text{RB}}^{\text{eff}}k_BT}, \label{eq:Branchallb} \\
n_{\text{FB}}^{\text{eff}} &= \left( n_{FB} + n_{FB} k_{FB} \frac{k_{B}T}{q} \ln\left( \frac{J}{J_{S_{FB}}[J]} +1\right)\right), \,\,\, and \label{eq:Branchallc} \\
V &= n_{\text{FB}}^{\text{eff}} \frac{k_{B}T}{q} \ln\left( \frac{J}{J_{S_{FB}}[J]}+1 \right) + (J - J_{D_{RB}}[J]) \times R_{RB}[J] \cdot A \label{eq:Branchalld}
\end{flalign}
\end{subequations}
\end{widetext}

In Eq. (\ref{eq:Branchall}), we present a comprehensive mathematical description of the current density (\(J\))-voltage (\(V\)) characteristics of a memristor, modeled as a series combination of a forward bias (FB) diode and a reverse bias (RB) diode. The effective ideality factors for the reverse and forward biased diodes are represented by \(n_{\text{RB}}^{\text{eff}}\) and \(n_{\text{FB}}^{\text{eff}}\), respectively, as shown in Eqs. (\ref{eq:Branchalla}) and (\ref{eq:Branchallc}). These effective ideality factors incorporate both the original ideality factors (\(n_{\text{RB}}\) and \(n_{\text{FB}}\)), the proprtionality factors (\(k_{\text{RB}}\) and \(k_{\text{FB}}\)), and the thermal voltage (\(k_BT/q\)), modified by the current density (\(J\)) and the saturation current densities (\(J_{S_{\text{RB}}}\) and \(J_{S_{\text{FB}}}\)) of the respective diodes similar to Eqn. \ref{eq:n_eff1}.

In equation (\ref{eq:Branchallb}), the current density through the reverse biased diode (\(J_{D_{\text{RB}}}\)) is expressed as a function of the applied voltage across the reverse biased diode (\(V_{\text{RB}}\)), the effective ideality factor (\(n_{\text{RB}}^{\text{eff}}\)), and the saturation current density (\(J_{S_{\text{RB}}}\)).

Finally, the overall voltage (\(V\)) across the memristor is described in equation (\ref{eq:Branchalld}) as a function of the current density (\(J\)), the effective ideality factor (\(n_{\text{FB}}^{\text{eff}}\)), the saturation current density (\(J_{S_{\text{FB}}}\)), the current density through the reverse biased diode (\(J_{D_{\text{RB}}}\)), the resistance of the reverse biased diode (\(R_{\text{RB}}\)), and the area of the top electrode (\(A\)).

Together, these equations provide a comprehensive mathematical framework with physical parameters for understanding and modeling the electrical characteristics of write and read states of memristors, represented as a series combination of forward and reverse biased diodes.

\begin{enumerate}
    \item \textbf{Write Branch 1 and 3:} The write branches are characterized by the redistribution of oxygen vacancies under applied write voltage. In write branch 1 oxygen vacancies are drifting from the \textit{TE} to the \textit{BE}. And in write branch 3 the oxygen vacancies are drifting from the \textit{BE} to the \textit{TE}. The redistribution of oxygen vacancies changes the alignment of the quasi-fermi level in \textit{BFO} with respect to the conduction band minimum and with that also the internal state parameter of the barrier height at the \textit{TE/BFO} and at the \textit{BFO/BE} interface. The other changing internal parameters are $n_{\text{eff}}$, $J_s$, and $R_s$ as a function of current density $J$ in equations \ref{eq:Branchall}.

    \item \textbf{Read Branch 2 and 4:} This voltage does not redistribute the oxygen vacancies as long as maximum amplitude of read voltage is not larger than the maximum voltage of previously applied write voltage. Distribution of oxygen vacancies will only occur if a voltage of opposite polarity in comparison to the previously applied write voltage is applied to the \textit{BFO} memristor. This again corresponds to the voltage ramping in a write branch. The current density at any voltage is accurately described by the internal state parameters, i.e., by the internal state variables at the largest bias of branch 1 for branch 2 and by the internal state variables at the largest bias of branch 3 for branch 4: $J_s$ and $R_s$. However, the ideality factor $n_{\text{FB}}^{\text{eff}}$ here is dependent on the voltage drop across the forward bias Schottky contact (eqn.\,\ref{eq:n_eff}).
\end{enumerate}

\section{\label{Meth}Methods}

The fitting procedures for branch 1-4, as delineated in Equation \ref{eq:Branchall}, were executed using Python in distinct methodologies to optimize accuracy and efficiency. Read branch 2, defined in Equation \ref{eq:Branchallb}, and read branch 4, as per Equation \ref{eq:Branchalld}, were subject to an automated fitting process. This approach leveraged the \texttt{least\_squares} function in Python, a robust optimization tool commonly employed in computational data analysis. The function iteratively minimized the residuals between the observed data and the model predictions, adjusting the parameters to best fit the empirical observations. Conversely, write branch 1 (Equation \ref{eq:Branchalla}) and write branch 3 (Equation \ref{eq:Branchallc}) necessitated an iterative manual fitting approach. Such a manual approach is particularly beneficial in cases where the model behavior is complex or non-linear, as it permits a more tailored fit to the dataset.

\section{Results}\label{res}

The interface switching in memristors is affected by the physical properties of the memristor material involved, and the relationships between the current density J and voltage V. The internal state variables of the memristor are crucial to understand its behavior and to develop electronic circuits with memristors, e.g. \textit{AI} accelerators. The presented equations in the appendix Eqns. \ref{eq:Branch1}, \ref{eq:Branch2}, \ref{eq:Branch3} and \ref{eq:Branch4} and the parameters listed in Tab.~\ref{tb:ParaTab} are related to the analysis of interface switching in three different \textit{BFO} memristors. 

The table provided (Table~\ref{tb:ParaTab}) presents a comprehensiveoverview of the physical model parameters pertaining to the three \textit{BFO} memristors (Samples 1-3) as they relate to four operational branches ($1\uparrow$, $2\downarrow$, $3\uparrow$, $4\downarrow$). These branches represent the intricate interactions occurring at the interfaces of the Top Electrode (\textit{TE})/\textit{BFO} and \textit{BFO}/Bottom Electrode (\textit{BE}) under varying bias conditions, encompassing both forward and reverse biases. It is worth highlighting that the parameters $n$ and $k$ employed in our modeling are intricately linked to the effective parameter $n_{\text{eff}}$, as elaborated in Eq.~\ref{eq:n_eff}.

\begin{table*}
\caption{Physical model parameters of samples 1-3 in branches 1-4, detailing the characteristics between the Top Electrode (\textit{TE})/\textit{BFO} and BFO/Bottom Electrode (\textit{BE}) interfaces in forward bias (FB) and reverse bias (RB). Modeled parameters \(n\) and \(k\) are associated with the \(n_{\text{eff}}\) as described in Eq.~\ref{eq:n_eff}.  The current density ($J$), reverse saturation current ($Js$) and $\frac{\partial \log(J_s)}{\partial \log(J)}$ is a proportionality constant are extracted from the modeled Eq.~\ref{eq:Branch1}-~\ref{eq:Branch4}  The leakage resistance ($R_s$) is multiplied with contact area (A = 4.53 $\times$ 10$^{-2}$ mm$^2$). The extracted parameters do not depend on the contact area. This representation is useful for memristors with interface switching.  Subscripts \textit{TE} and \textit{BE} are inserted for $n$, $k$, $J_s$, $R_s$ and $\frac{\partial \log(J_s)}{\partial \log(J)}$ to represent top electrode and bottom electrode parameters.}
\label{tb:ParaTab}
\begin{centering}

\begin{tabular}{c|cccc|cc|cccc|cc} 
  \hline 
   \hline \\
 &\multicolumn{4}{c}{\textbf{TE/BFO}} & \multicolumn{2}{c}{\textbf{BFO/BE}} &\multicolumn{4}{c}{\textbf{BFO/BE}} & \multicolumn{2}{c}{\textbf{TE/BFO} } \\
 
 &\multicolumn{4}{c}{\textit{(1$\uparrow$ \& 2$\downarrow$ $\rightarrow$ \textit{TE}:\textit{FB}, 3$\uparrow$ $\rightarrow$ \textit{TE}:\textit{RB})}} & \multicolumn{2}{c}{ \textit{(1$\uparrow$ \& 2$\downarrow$ $\rightarrow$ \textit{BE}:\textit{RB})}} &\multicolumn{4}{c}{\textit{($\uparrow$3 \& 4$\downarrow$ $\rightarrow$ \textit{BE}:\textit{FB}, 1$\uparrow$ $\rightarrow$ \textit{BE}:\textit{RB})}} & \multicolumn{2}{c}{\textit{(3$\uparrow$ \& 4$\downarrow$ $\rightarrow$ \textit{TE}:\textit{RB})}} \\
 
\text{Branch} & $n_{TE}$ & $k_{TE}$ & $Js_{TE}$ &
\multirow{2}{*}{$\frac{\partial \log(Js_{TE})}{\partial \log(J)}$} & $Rs_{BE}.A$ &
\multirow{2}{*}{$\frac{\partial \log(Rs_{BE})}{\partial \log(J)}$} & $n_{BE}$ & $k_{BE}$ & $Js_{BE}$ &
\multirow{2}{*}{$\frac{\partial \log(Js_{BE})}{\partial \log(J)}$} & $Rs_{TE}.A$ &
\multirow{2}{*}{$\frac{\partial \log(Rs_{TE})}{\partial \log(J)}$}\\
& &V$^{-1}$ &$nA/mm^{2}$& & k$\Omega \cdot mm^{2}$ & & &V$^{-1}$ &$nA/mm^{2}$ & & k$\Omega \cdot mm^{2}$ & \\
 \hline
 \multicolumn{13}{c}{\text{Sample 1}}\\ 

1$\uparrow$	& 32.00 & 0.30 & 540.84 & +0.45 & 21.34 & -0.81 & 4.88 & 0.01 & 234.00 & - & - & -\\ 
2$\downarrow$	& 24.00 & 2.75 & 540.84 & - & 21.34 & - & - & - & - & - & - & -\\ 
3$\uparrow$	& 24.00 & 60.00 & 540.84 & - & - & - & 23.00 & 0.40& 234.00 & +0.40 & 2.58 & -0.99\\ 
4$\downarrow$	& - & - & - & - & - & - & 4.88 & 19.6 & 234.00 & - & 2.58 & -\\ 
 \hline
 \multicolumn{13}{c}{\text{Sample 2}}\\ 

1$\uparrow$	& 16.00 & 0.20 & 624.72 & +0.65 & 3.64 & -0.80 & 11.40 & 6500.00 & 206.62 & - & - & -\\ 
2$\downarrow$	& 15.80 & 0.33 & 624.72 & - & 3.64 & - & - & - & - & - & - & -\\ 
3$\uparrow$	& 15.80 & 4000.00 & 624.72 & - & - & - & 16.00 & 0.15 & 206.62 & +0.20 & 25.55 & -0.93\\ 
4$\downarrow$	& - & - & - & - & - & - & 11.40 & 2.42 & 206.62 & - & 25.55 & -\\ 
 \hline
 \multicolumn{13}{c}{\text{Sample 3}}\\ 

1$\uparrow$	& 19.00 & 0.16 & 103.09 & +0.30 & 2.17 & -0.90 & 6.85 & 0.1 & 7.04 & - & - & -\\ 
2$\downarrow$	& 9.24 & 4.50 & 103.09 & - & 2.17 & - & - & - & - & - & - & -\\ 
3$\uparrow$	& 9.24 & 1200.00 & 103.09 & - & - & - & 19.00 & 3.30 & 7.04 & +0.25 & 126.84 & -1.10\\ 
4$\downarrow$	& - & - & - & - & - & - & 6.85 & 18.10 & 7.04 & - & 126.84 & -\\
 \hline
  \hline 
\end{tabular}

\end{centering}

\end{table*}

\begin{figure*}
  \centering
\includegraphics[width=\textwidth]{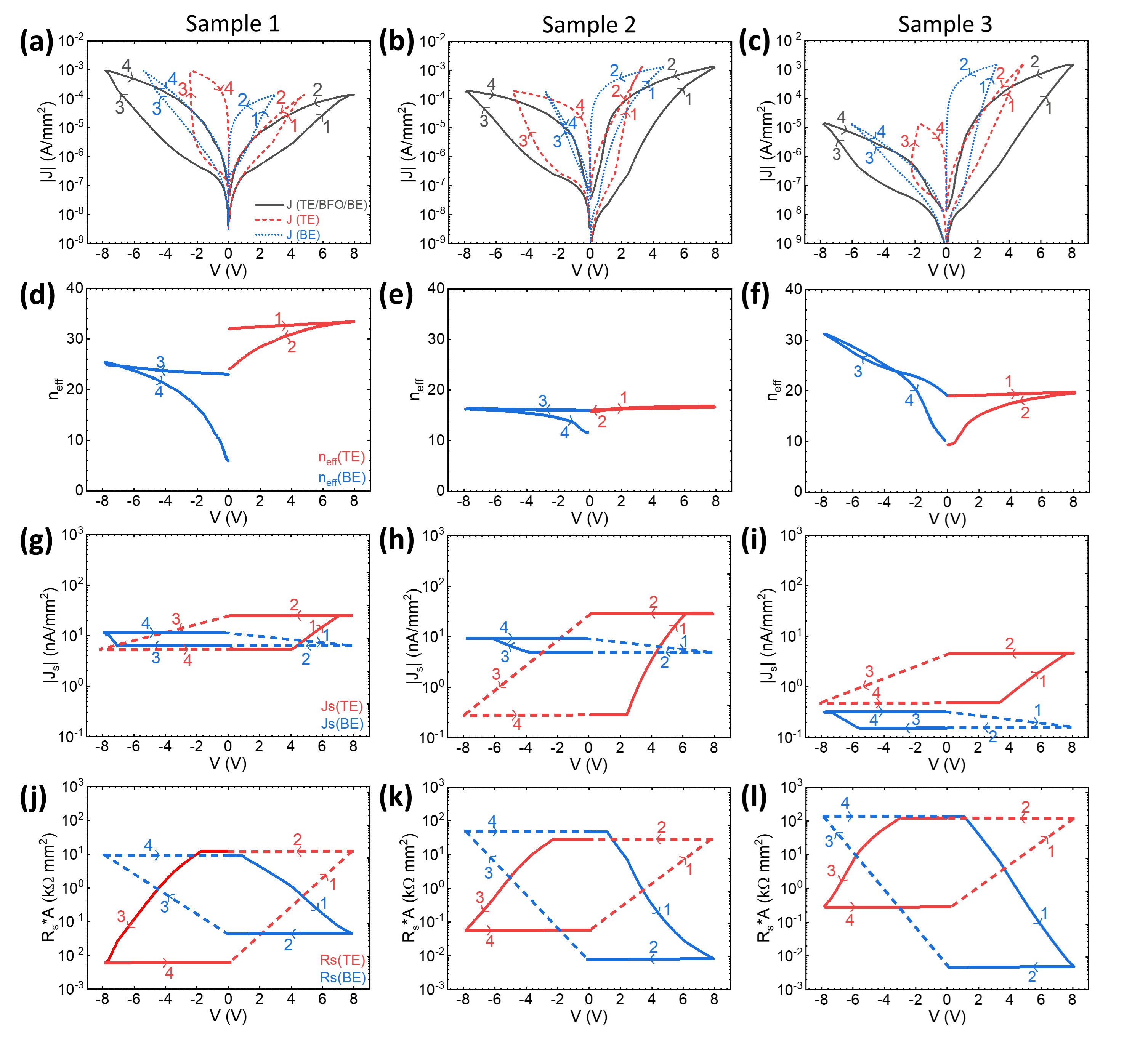}
\caption{\label{fig:Mresul} (a,b,c) Modeled current density versus voltage (J-V) curves for the \textit{BFO} memristor with two depletion regions (\textit{TE/BFO/BE}), top electrode Schottky contact (\textit{TE/BFO}), and the bottom electrode Schottky contact (\textit{BFO/BE}). The modeled parameters  (d,e,f) \(n_{\text{eff}}\), (g,h,i) \(J_{\text{S}}\), and  (j,k,l) \(R_{\text{S}}\) are shown as a function of voltage for samples 1-3. The physical model parameters encompass the ideality factor (\(n_{\text{eff}}\)), the reverse saturation current (\(J_{\text{S}}\)), and the leakage resistance (\(R_{\text{S}}\)). Extracted parameters for \textit{TE} and \textit{BE} are shown in red and blue color, respectively.}
\end{figure*}

Specifically, within branches $1\uparrow$ and $2\downarrow$, the \textit{TE/BFO} interface operates as a forward bias diode (\textit{FB}), while the \textit{BFO/BE} interface functions as a reverse bias diode (\textit{RB}) (Refer to Eqns. (\ref{eq:Branch1})-(\ref{eq:Branch4})). Conversely, in branches $3\uparrow$ and $4\downarrow$, this configuration is reversed, with the \textit{TE/BFO} acting as \textit{RB} and \textit{BFO/BE} as \textit{FB} diodes. In the context of branches $2\downarrow$ and $4\downarrow$, it is noteworthy that the parameters $n$, $k$, $J_s$, and $R_s$ exhibit no discernible variation in response to applied voltage changes. These parameters remain constant, underscoring their stability in these specific operational conditions. In contrast, within branches $1\uparrow$ and $3\uparrow$, we observe that the parameters $n$, $k$, and $J_s$ for the RB diode remain unchanged over small voltage ranges, typically below 2\,V. Furthermore, above 2\,V, any variation in these parameters with applied voltage is deemed negligible for our modeling purposes. Notably, the parameters $n$ and $k$ for the \textit{FB} diode hold validity across all voltage ranges in these branches and remain unaltered by changes in applied voltage.

It is crucial to emphasize that the voltage-dependent parameters in these branches are $J_s$ (\textit{FB}) and $R_s$ (\textit{RB}). We observe that the logarithm of $J_s$ is directly proportional to the logarithm of $J$, while the logarithm of $R_s$ is directly proportional to the negative logarithm of $J$. The precise proportionality constants governing these relationships are meticulously documented in Table~\ref{tb:ParaTab}, as $\frac{\partial \log(Js_{TE or BE})}{\partial \log(J)}$ and $\frac{\partial \log(Rs_{TE or BE})}{\partial \log(J)}$, respectively. For a given cycle of resistive switching (\textit{RS}) in \textit{2D} memristors, which exhibit significant cycle-to-cycle (\textit{C2C}) variability, Spetzler et al. \cite{spetzler_reference} also determined a proportionality constant between $\log I_s$ and $\log I$ \cite{spetzler_reference}. Given the negligible \textit{C2C} variability in \textit{BFO} memristors, all parameters in Table~\ref{tb:ParaTab}, including the proportionality constant between $\log I_s$ and $\log I$, are applicable for all cycles of \textit{RS} in \textit{BFO} memristors. It is noteworthy that as voltage increases, the barrier height of the \textit{FB} diode decreases ($J_s$ increases), while the leakage resistance $R_s$ of the RB diode decreases correspondingly. Consequently, there exists an inverse relationship between the leakage resistance $R_s$ of the \textit{RB} diode and the barrier height of the \textit{RB} diode, with both parameters exhibiting voltage-dependent behaviors.

Furthermore, our investigation revealed intriguing insights into the behavior of these diodes as a function of voltage. The order of change in the resistance parameter, denoted as $R_s$, of the reverse bias (\textit{RB}) diode is directly correlated with the maximum electrical current ($J$) at both the maximum positive voltage, denoted as $+V_{\text{max}}$, and the maximum negative voltage, denoted as $-V_{\text{max}}$. This relationship can be exemplified by comparing our findings for Samples 1, 2, and 3, as illustrated in Figure \ref{fig:EnM} and documented in Table~\ref{tb:ParaTab}. For instance, at $+V_{\text{max}}$, we observed $R_s$ values of $21.34 \, k\Omega \cdot \text{mm}^{2}$, $3.64 \, k\Omega \cdot \text{mm}^{2}$, and $2.17 \,k\Omega \cdot \text{mm}^{2}$ for Samples 1, 2, and 3, respectively, while the corresponding maximum $J$ values were 1.35 $\times 10^{-4}$\,$A/mm^2$, 1.24 $\times 10^{-3}$\,$A/mm^2$, and 1.46 $\times 10^{-3}$\,$A/mm^2$. Similarly, at $-V_{\text{max}}$, $R_s$ values were $2.58 \, k\Omega \cdot \text{mm}^{2}$, $25.55 \, k\Omega \cdot \text{mm}^{2}$, and $126.84 \, k\Omega \cdot \text{mm}^{2}$ for the same samples 1, 2, and 3, respectively, with corresponding maximum absolute $J$ values of 8.74 $\times 10^{-3}$\,$A/mm^2$, 1.75 $\times 10^{-4}$\,$A/mm^2$, and 1.38 $\times 10^{-5}$\,$A/mm^2$.

Figure \ref{fig:Mresul} shows the extracted parameters from Eqns.~\ref{eq:Branch1}--\ref{eq:Branch4}. The extracted parameters for \textit{TE} and \textit{BE} are shown in red and blue color, respectively. Figure \ref{fig:Mresul} (a,b,c) shows simulated voltage drop over sample (\textit{TE/BFO/BE}), and over junctions (\textit{TE/BFO} and \textit{BFO/BE}) for current density \(J\). The effective ideality factor parameter, denoted as \(n_{\text{eff}}\) (Eq.~\ref{eq:n_eff}) and \(J_s\), at the interfaces of \textit{TE/BFO} and \textit{BFO/BE} during the transition between operational branches ($1\uparrow \rightarrow 2\downarrow \rightarrow 3\uparrow \rightarrow 4\downarrow \rightarrow 1\uparrow$), is depicted in Fig. \ref{fig:Mresul} (d,e,f)  and in Fig. \ref{fig:Mresul} (g,h,i), respectively. The \(n_{\text{eff}}\) and \(J_s\) are parameters that influence one another as they belong to the same diode. Similarly, for reverse bias (\textit{RB}) diodes, the parameter \(R_s\) is represented by solid lines in Figure \ref{fig:Mresul} (j,k,l). It is important to note that the dotted lines in the figure represent estimated variations in an unknown branch, and these estimates may not be perfectly accurate. \(J_s\) typically changes only above \(2\,V\) and \(R_s\) also generally changes above \(1\,V\), signifying the influence of the oxygen vacancy drift only after a threshold has been reached. As illustrated in Figure~\ref{fig:Mresul}\,(g,h,i), the saturation current density \( J_s \) exhibits a plateau beyond a certain high voltage threshold, indicating a stabilization in its value despite further voltage increases. This behavior is in agreement with the intermediate stages presented in Figure~\ref{fig:bandE}, where the ferroelectric diode barrier height \( \Phi{}B \) has reached its minimum level, particularly at the \textit{BFO/BE} interface in Branch 3, and the \textit{TE/BFO} interface in Branch 1. These insights are pivotal for simulating electronic circuits with interface-type memristors tailored for particular applications.

In conclusion, the analysis of Table~\ref{tb:ParaTab} and figure \ref{fig:Mresul} reveals crucial characteristics of the three distinct memristor samples across the four different operation branches 1-4. These insights are pivotal for optimizing memristor performance and can inform the design of memristors with specific characteristics tailored for particular applications. 

\section{Conclusions and Outlook}\label{CnO}

A physical memristor model of hysteretic current-voltage curves of the interface-type \textit{BFO} memristor with hysteretic current-voltage curves has been developed. It assumes a change of internal state variables in the two write branches due to dominating ion drift and no change of the internal state variables in the two read branches where electron drift dominates. Primarily, the interface-type \textit{BFO} memristor has been considered as a system with two Schottky diodes in series. Every Schottky diode, the Schottky diode forming a depletion layer at the bottom electrode (BE) and the Schottky diode forming a depletion layer at the top electrode (TE), is described by an ideal Schottky barrier and a leakage resistance in parallel. The Schottky barrier and the leakage resistance of the reverse/forward Schottky diode at the \textit{TE} and of the forward/reverse Schottky diode at the \textit{BE} have been analyzed. Reported changes of barrier heights and leakage resistance correctly describe the current flow through the reverse/forward Schottky diode at the \textit{TE} and the forward/reverse Schottky diode at the \textit{BE}. Our findings rectify a previously misconstrued model of the Schottky barriers in interface-type, analog memristors, highlighting that both electrodes possess barriers, albeit one being dominantly shunted by the parallel leakage resistance. The redistribution of oxygen vacancies in the $n - \text{BiFeO}_3$ (BFO) layer and its effect on the position-dependent band alignment of the conduction band minimum and of the quasi-Fermi level in \textit{BFO} have been meticulously analyzed. It is the difference between $E_C$ and $E_F$ at the \textit{TE}/\textit{BFO} interface which influences the barrier height of the Schottky barrier at the \textit{TE} and it is the difference between $E_C$ and $E_F$ at the \textit{BFO}/\textit{BE} interface which influences the barrier height of the Schottky barrier at the \textit{BE}. A redistribution of oxygen vacancies will only occur and affect $E_C$ and $E_F$ and leakage resistance in the write branches 1 and 3. Our analysis has shed light on the correct dependence of Schottky barrier height on the branch type being either write branch 1 and branch 3 or read branch 2 and branch 4. This marks a significant stride towards a physical memristor model describing the internal state variables at every point of the hysteretic current-voltage curves of interface-type, analog memristors.

The insights garnered from the presented study not only contribute to the burgeoning literature on memristor technology, but also pave a promising avenue for the use of such interface-type, analog, and non-linear memristors in electronic circuits for \textit{AI} accelerators. Using the presented physical memristor model for electronic design automation simulations, the power consumption, speed, size, and cost of the electronic circuit can be estimated and compared with standard \textit{CMOS} solutions.\\

\begin{acknowledgments}
We wish to thank the Bundesagentur für Bildung und Forschung for financial support (BMBF project ForMikro - ERMI 16ES1119).
\end{acknowledgments}

\appendix

\section{J-V Characteristic Equation for All Branches}\label{secA1}

In equations (\ref{eq:Branch1})-(\ref{eq:Branch4}), we present a detailed mathematical framework for different branches of a memristor's J-V characteristics. Each branch corresponds to a different operational state of the memristor: write or read, as indicated by the up arrow (\(\uparrow\)) for write branches and down arrow (\(\downarrow\)) for read branches, and each branch has a unique set of parameters.

For Branch 1 (Eq. \ref{eq:Branch1}), the equations describe the operation during a write operation (indicated by the superscript \(1\uparrow\)) where \(n_{\text{{BE}}}^{\text{{eff}}}\) and \(n_{\text{{TE}}}^{\text{{eff}}}\) are the effective ideality factors for the \textit{BE} and \textit{TE} diodes, respectively, and \(J_{D_{BE}}\) is the current density through the \textit{BE} diode. \(V\) is expressed as a function of the current density \(J\), effective ideality factors, and other parameters.\\
Branch 2 (Eq. \ref{eq:Branch2}) describes the operation during a read operation (indicated by the superscript \(2\downarrow\)), where \(n_{\text{{eff}}}^{TE}\) is the effective ideality factor for the \textit{TE} diode, and \(V\) is calculated based on the current density \(J\), \(n_{\text{{eff}}}^{TE}\), and other parameters.
Following are equations for read (2 and 4) and write (1 and 3) branches.
\begin{widetext}
\begin{subequations}
\label{eq:Branch1}
\begin{flalign}
\text{Branch 1:} & \nonumber \\
n_{\text{{BE}}}^{\text{{eff}}} & = \left( n_{BE}^{1\uparrow} + n_{BE}^{1\uparrow} k_{BE}^{1\uparrow} \frac{k_{B}T}{q} \ln\left( \frac{-J_{D_{BE}}}{J_{S_{BE}}[-J]}+1 \right)\right), && \label{eq:Branch1a} \\
J_{D_{BE}} & = J_{S_{BE}}[J] \left( e^{\Delta} - 1 \right)\,\,\, where \,\,\, \Delta = \frac{-qV_{BE}}{n_{\text{{BE}}}^{\text{{eff}}}k_BT}, && \label{eq:Branch1b} \\
n_{\text{{TE}}}^{\text{{eff}}} & = \left( n_{TE}^{1\uparrow} + n_{TE}^{1\uparrow} k_{TE}^{1\uparrow} \frac{k_{B}T}{q} \ln\left( \frac{J}{J_{S_{TE}}[J]} +1\right)\right),\,\,\, and && \label{eq:Branch1c}\\
V & = n_{\text{{TE}}}^{\text{{eff}}} \times \frac{k_{B}T}{q} \ln\left( \frac{J}{J_{S_{TE}}[J]}+1 \right) + (J - J_{D_{BE}}[J]) \times R_{BE}[J] \cdot A && \label{eq:Branch1d}
\end{flalign}
\end{subequations}
 
\begin{subequations}
\label{eq:Branch2}
\begin{flalign}
\text{Branch 2:} & \nonumber \\
n_{\text{{eff}}}^{TE} & = \left( n_{TE}^{2\downarrow} + n_{TE}^{2\downarrow} k_{TE}^{2\downarrow} \frac{k_{B}T}{q} \ln\left( \frac{J}{I_{S_{TE}}[+J_{\text{{max}}}]}\,+1 \right)\right) \,\,\, and && \label{eq:Branch2a} \\
V & = n_{\text{{eff}}}^{TE} \times \frac{k_{B}T}{q} \ln\left( \frac{J}{J_{S_{TE}}[+J_{\text{{max}}}]}\,+1 \right) + J \times R_{BE}[+J_{\text{{max}}}] \cdot A   && \label{eq:Branch2b}
\end{flalign}
\end{subequations}

\begin{subequations}
\label{eq:Branch3}
\begin{flalign}
\text{Branch 3:} & \nonumber \\
n_{\text{{TE}}}^{\text{{eff}}} & = \left( n_{TE}^{3\uparrow} + n_{TE}^{3\uparrow} k_{TE}^{3\uparrow} \frac{k_{B}T}{q} \ln\left( \frac{-J_{D_{TE}}}{J_{S_{TE}}[J]}+1 \right)\right), && \label{eq:Branch3a} \\
J_{D_{TE}} & = J_{S_{TE}}[J] \left( e^{\Delta} - 1 \right) \,\,\, where \,\,\, \Delta = \frac{-qV_{TE}}{n_{\text{{TE}}}^{\text{{eff}}}k_BT}, && \label{eq:Branch3b} \\
n_{\text{{BE}}}^{\text{{eff}}} & = \left( n_{BE}^{3\uparrow} + n_{BE}^{3\uparrow} k_{BE}^{3\uparrow} \frac{k_{B}T}{q} \ln\left( \frac{J}{J_{S_{BE}}[J]} +1\right)\right)\,\,\, and && \label{eq:Branch3c}\\
V & = n_{\text{{BE}}}^{\text{{eff}}} \times \frac{k_{B}T}{q} \ln\left( \frac{J}{J_{S_{BE}}[J]}+1 \right) + (J - J_{D_{TE}}[J]) \times R_{TE}[J] \cdot A  && \label{eq:Branch3d}
\end{flalign}
\end{subequations}

\begin{subequations}
\label{eq:Branch4}
\begin{flalign}
\text{Branch 4:} & \nonumber \\
n_{\text{{eff}}}^{BE} & = \left( n_{BE}^{4\downarrow} + n_{BE}^{4\downarrow} k_{BE}^{4\downarrow} \frac{k_{B}T}{q} \ln\left( \frac{-J}{I_{S_{BE}}[-J_{\text{{max}}}]}\,+1 \right)\right)\,\,\, and && \label{eq:Branch4a} \\
V & = -n_{\text{{eff}}}^{BE} \times \frac{k_{B}T}{q} \ln\left( \frac{-J}{J_{S_{BE}}[-J_{\text{{max}}}]}\,+1 \right) + J \times R_{TE}[-J_{\text{{max}}}] \cdot A && \label{eq:Branch4b}
\end{flalign}
\end{subequations}
\end{widetext}
For Branch 3 (Eq. \ref{eq:Branch3}), similar to Branch 1 but for a different segment of the write operation (indicated by the superscript \(3\uparrow\)), \(n_{\text{{TE}}}^{\text{{eff}}}\) and \(n_{\text{{BE}}}^{\text{{eff}}}\) are the effective ideality factors for the \textit{TE} and \textit{BE} diodes, respectively, and \(J_{D_{TE}}\) is the current density through the \textit{TE} diode.

Finally, Branch 4 (Eq. \ref{eq:Branch4}) corresponds to another read operation (indicated by the superscript \(4\downarrow\)), where \(n_{\text{{eff}}}^{BE}\) is the effective ideality factor for the \textit{BE} diode, and \(V\) is calculated based on the current density \(J\), \(n_{\text{{eff}}}^{BE}\), and other parameters.

Together, these equations describe the memristor's I-V characteristics for different operational states (write or read) and enable a detailed understanding and modeling of the memristor's behavior.

\bibliography{bib}

\end{document}